\documentclass[a4paper]{spie}  
\usepackage[]{graphicx}

\title{Design and construction of the IMACS-IFU, a 2000-element integral field unit} 

\author{J\"urgen Schmoll\supit{a}, George N. Dodsworth\supit{a}, Robert Content\supit{a} and Jeremy R. Allington-Smith\supit{a} 
\skiplinehalf
\supit{a}Astronomy Instrument Group, Dept. of Physics, Rochester Building, University of Durham, South Road, Durham DH1 3LE, United Kingdom \\
}

\authorinfo{Further author information: (Send correspondence to J.S.) \\
J.S.: E-mail: jurgen.schmoll@durham.ac.uk, Telephone: 0044-(0)191-334-3541 \\ }

\begin{document} 
  \maketitle 

\begin{abstract}
The IMACS-IFU is an Integral Field Unit built for the IMACS spectrograph at the Magellan-I-Telescope at Las Campanas Observatory. It consists of two rectangular fields of 5 by 7 square arc seconds, separated by roughly one arc minute. With a total number of 2000 spatial elements it is the second largest fiber-lenslet based IFU worldwide, working in a wavelength range between 400 and 900 nm. Due to the equally sized fields classical background subtraction, beam switching and shuffling are possible observation techniques. One particular design challenge was the single, half a metre long curved slit in combination with a non telecentric output. Besides the construction some preliminary results are described.
\end{abstract}



\section{INTRODUCTION}

Apart from bare lenslet approaches as SAURON (Bacon et al, 2001 \cite{Bacon2001}) and the image slicer principle (Content 1998 \cite{Content1998}), the fiber-lenslet coupled integral field units are often used to sample the sky. Depending on the scientific goals, the number of elements can be either quite small to leave enough detector space for each spectrum for high spectrophotometric accuracy as in the PMAS case (Kelz et al, 2003 \cite{Kelz2003}), or a higher number of elements can be used to get a finer sampling or a larger field of view as for TEIFU (Murray et al, 2000 \cite{Murray2000}). For background subtraction purposes a second field may be introduced as it has been done for the GMOS-IFU (Allington-Smith et al, 2002 \cite{Allington-Smith2002b}). The IMACS IFU described here belongs to the latter ones, having two fields of similar size to enable beam switching by nodding the telescope. Each field consists of 1000 elements to get a field of view of 5 by 7 arc seconds, combined with a fine sampling of 0''.2. Behind the VIMOS-IFU (LeFevre et al, 2000 \cite{LeFevre2000}) it is the second largest operational fiber based IFU existing.

\label{sect:intro}  


\section{REQUIREMENTS AND SPECIFICATION} 
\label{sect:requirements}
The IMACS spectrograph has been designed for imaging, long-slit and multi-slit spectroscopy on a field of view being 27 arc minutes in diameter. The whole setup is located on the Nasmyth platform and mechanically derotated. The basic parameters are shown in table \ref{basicparameters} (Dressler 1999 \cite{Dressler1999}). The spectrograph has two cameras (called long and short camera in the following). While the short camera works with grisms of rather low dispersion, the long camera is accessed by a reflective grating. Alternatively both cameras can be used for imaging when the dispersing element is removed (short camera case) or replaced by a plane mirror (long camera case).

\begin{table}[h]
\begin{center}
\begin{small}
\begin{tabular}{|l|l|l|l|}
\hline
Instrument & Mode & Property & Value \\
\hline
TELESCOPE & Nasmyth & Aperture & 6.5 m \\
          &         & Focal length & 71.5 m \\
          &         & Focal ratio & f/11 \\
          &         & System & Gregorian \\
          &         & Additional items & ADC, field corrector at M3 \\
          &         & Field curvature & R = 1231.74 mm \\
\hline
IMACS IMAGER & LONG camera f/2.7  & Field size & 15'.15 $\times$ 15'.15 \\
             &                    & Resolution & 0''.111 / pixel \\
             & SHORT camera f/1.5 & Field size & 27.3' $\times$ 27.3' \\
             &                    & Resolution & 0''.2 / pixel \\
\hline
IMACS SPECTROGRAPH & LONG camera f/2.7  & long slit length & 15' \\
                   &                    & spectral range & [365..1000] nm \\  
                   & SHORT camera f/1.5 & long slit length & 27' \\
                   &                    & spectral range & [390..1050] nm \\
\hline
\end{tabular}
\caption{MAGELLAN Telescope and IMACS instrument parameters}
\label{basicparameters}
\end{small}
\end{center}
\end{table}

\subsection{Telescope and Spectrograph interfaces}
The IMACS spectrograph has been designed mainly with long or multislit use in mind. For this curved masks can be fed via a slit mask changer into the focal plane. Due to the strong field curvature these masks have to be bent into a spherical shape. The IFU uses the same mask interface, while it needs the space of three masks in the storage cassette. While the two fields of interest are accessed by two pickoff mirrors, the output consists of a single, long and curved slit. Hence the output microlenses were not only necessary to feed the f/11 collimator but also to steer the beam into the pupil. A staircase-like approximation was used to realize the slit curvature. The mass for the assembly had to be restricted to 10 kg to be compatible with the mask changer mechanism.

\subsection{Requirements}
The summarized parameters and requirements for the IFU are
\begin{itemize}
\item Wavelength range 400 to 900 nm
\item Efficiency 50 \% or higher
\item Two fields, separated by fixed 60'' or 20.8 mm in the telescopes focal plane
\item 25 $\times$ 40 elements in a matrix with full spatial coverage
\item Spatial sampling 0.2 arcsec per element \\
$\rightarrow$ sky coverage 6''.92 $\times$ 5''.00 (short camera) or 4''.15 $\times$ 5''.00 (long camera).
\item Both fields of same size to simplify beam switching
\item Ratio of field sidelengthes $\approx 1 / \sqrt{2}$ to preserve aspect ratio when mosaicing
\item Curved output pseudoslit matched to f/11 IMACS collimator and non telecentric collimator incidence
\item Maximum weight 10 kilograms
\item Passive element without moving parts inside
\item To be used with both spectrograph cameras without change of field centers
\item Space between adjacent spectra 3.54 pixels (short camera) or 6.37 pixels (long camera)
\end{itemize}

Due to the wavelength range being mostly in the visual an optical fiber design was chosen. The design and manufacture could benefit from fiber-lenslet coupled IFUs being built in Durham before, namely TEIFU (Haynes et al, 1998 \cite{Haynes1998}, Murray et al, 2000 \cite{Murray2000}) and the GMOS-IFU (Allington-Smith et al, 2002 \cite{Allington-Smith2002}). As in these IFUs the input array consists of hexagonal elements of 400 $\mu$m diameter and linear arrays of truncated circular shape on the output side.


\section{OPTICAL DESIGN}
The optical design consists of two parts: The fore optics and the fiber-lenslet coupling. While the fore optics calculation is straightforward, the fiber-lenslet coupling required a more enhanced optimization process.
\label{sect:opticaldesign}

\begin{figure}[h]
\begin{center}
\begin{tabular}{c}
\includegraphics[height=10cm]{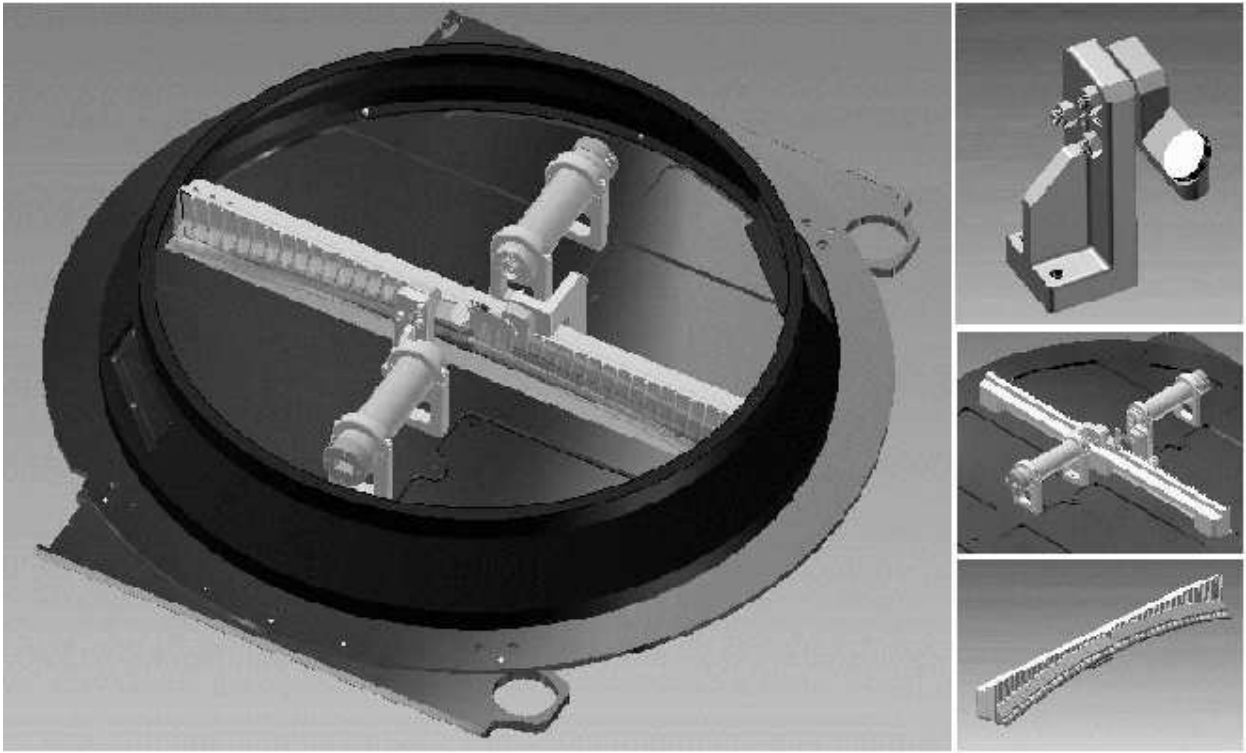}
\end{tabular}
\end{center}
\caption{The key assembly parts of the IFU and their arrangement. Right column: Above pickoff mirror mount, center all essential parts mounted on base plate, below the curved slit assembly. The diameter of the round base plate is 686 mm, while the slit is 526 mm long. \label{fig:imacsexplained_hori}}
\vspace{5mm}
\end{figure}

\begin{figure}[h]
\begin{center}
\begin{tabular}{cc}
\includegraphics[height=6cm]{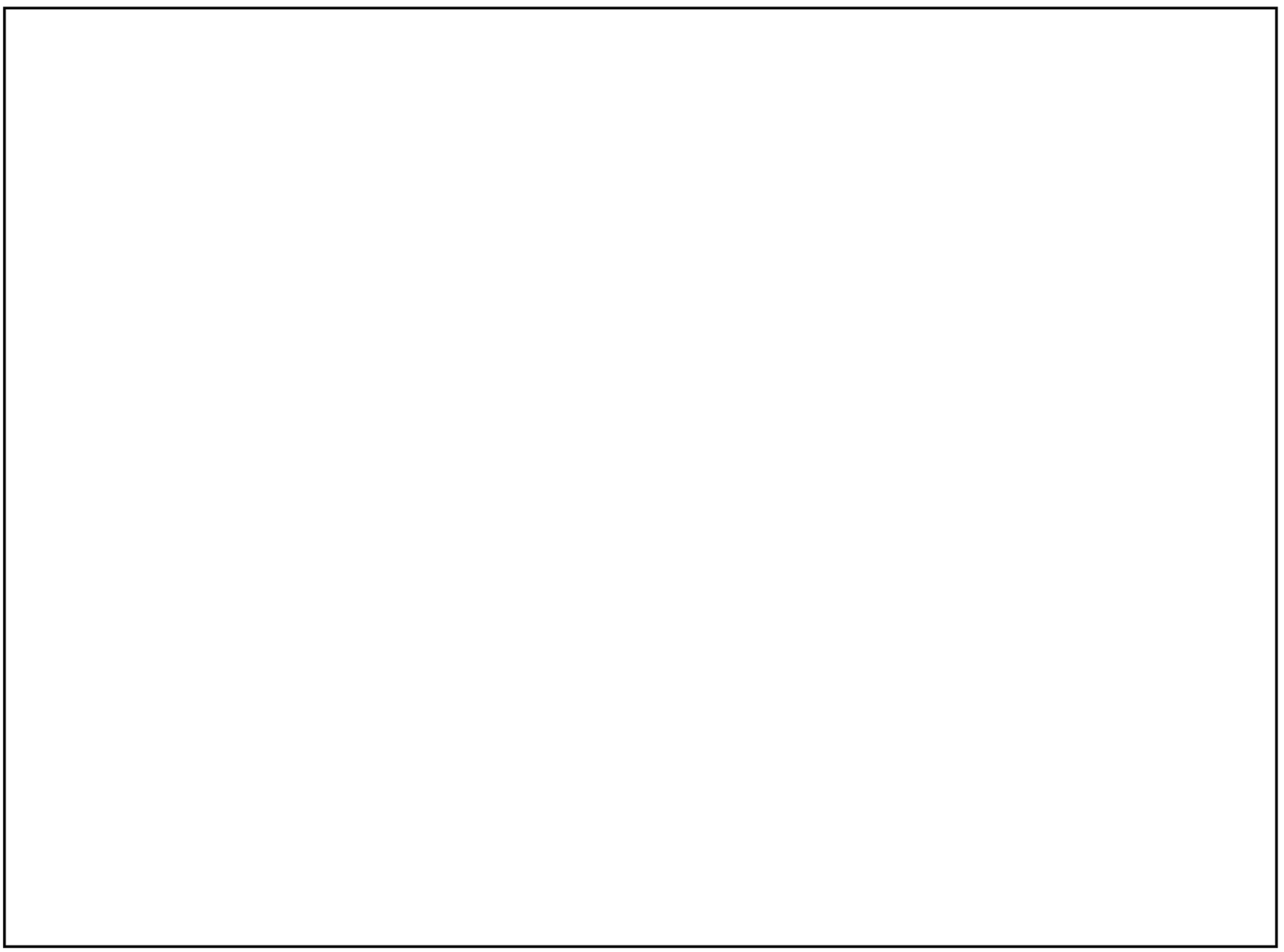}
\includegraphics[height=6cm]{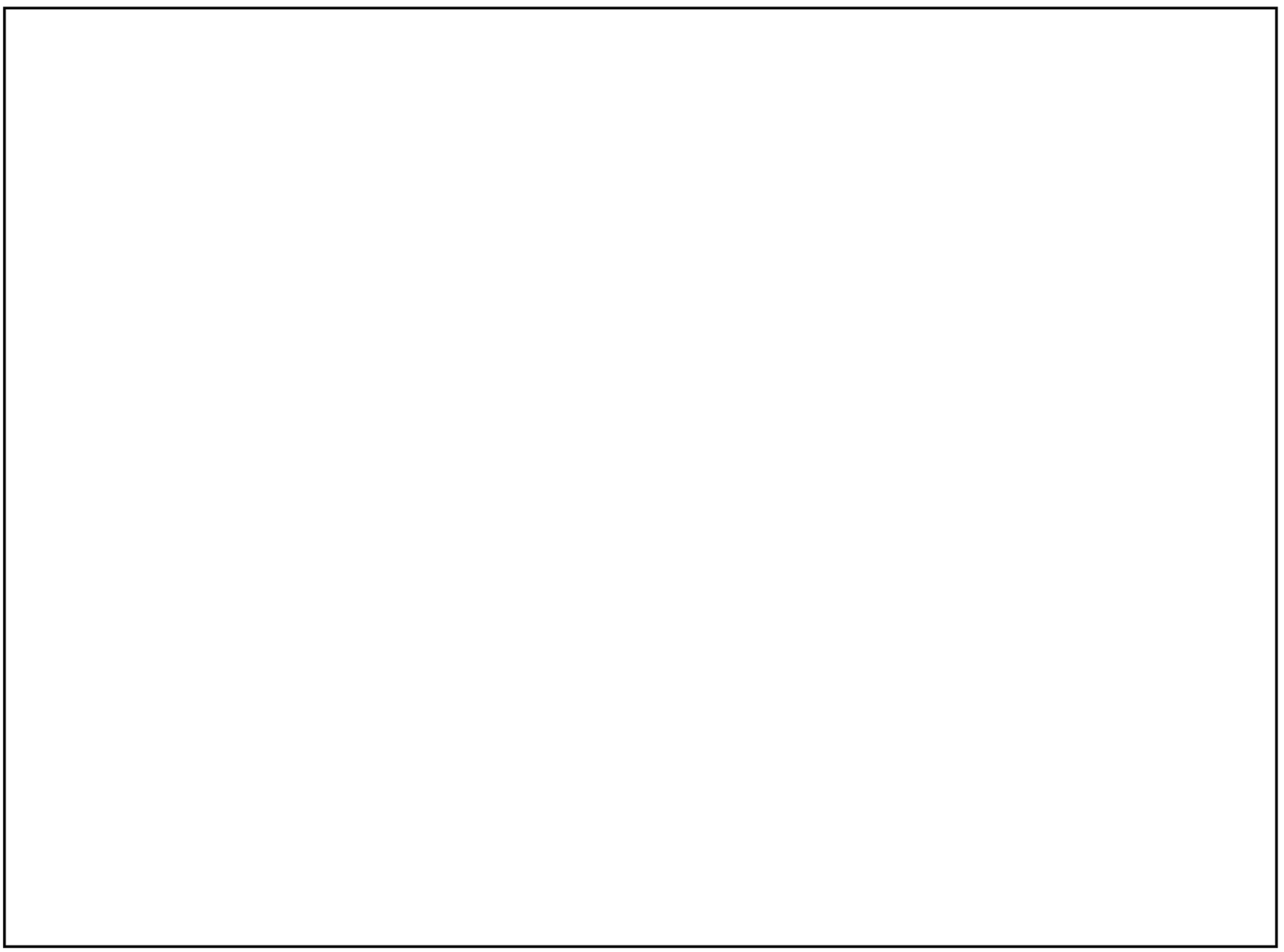}
\end{tabular}
\end{center}
\caption{The fore optics consists of two doubletts of BK7 and SF5, being 8 mm and 22 mm in diameter. Right: Spot diagrams, the scale bar is 1 mm and the dark circle the diffraction limit. \label{fig:foreoptics}}
\end{figure}

\subsection{Fore optics}
The fore optics system is required for two reasons. At first it changes the image scale to create the spatial sampling required for the IFU input. To match 0.2 arcseconds to 400 $\mu$m lenslets a magnification of 5.77 is required. The second requirement is more striking for the efficiency of the IFU: The fore optics has to assure a highly telecentric input up to the edges of the array. Hence each microlens focus is positioned at the core of the corresponding fiber. The system used consists of two all-spherical doubletts of BK7 and SF5. Figure \ref{fig:foreoptics} shows the design chosen. The spot diagrams are provided for the paraxial case (upper left), the half-field case (upper right) and the field corner (below). They show that the image is not diffraction limited, but about 90 \% of the spots energy matches a single input lenset.

\subsection{Fiber-Lenslet coupling}

The fiber-lenslet coupling assures that
\begin{itemize}
\item{a maximum fraction of light hitting the lenslet array is focused into the fiber cores}
\item{the fiber is fed with a fast focal ratio to minimize FRD losses}
\item{a maximum fraction of light leaving the fiber is inside the collimator f/11 cone}
\end{itemize}

To maximize the IFU efficiency an optimization of the global system has been undertaken. Parameters have been focal length and substrate thickness of in- and output arrays, core diameter and FRD properties of the fiber. Also the output focal ratio had to be chosen to take lateral displacement of the output pupil due to the beam steering limitation of the block design into account. These issues are discussed in section \ref{sect:curvslit}. The epoxy lenslet arrays on a BK7 substrate have been produced by AOA Inc. \footnote{Adaptive Optics Associates Inc., Cambridge MA, USA}. The fibers are mounted into linear arrays of 50 fibers each, held in place by silica v-grooves and epoxy. At the input the fibers have been fed into a tube stack to create the required two-dimensional pitch of the input lenslet array. Input and output blocks have been ground and optically polished before the micro lens arrays have been glued in place after alignment. Using UV curing glue which has a refractive index similar to the fibers and the substrate, not only the Fresnel losses disappear but also the effective FRD is reduced by minimising the effects of the fiber end faces (Schmoll et al, 2002 \cite{Schmoll2003}).

\subsection{Curved slit and non-telecentric output}
\label{sect:curvslit}

Due to the curved focal plane of the Magellan telescopes two more complications had been to solve. Apart from the defocus of up to about 20 mm when a straight slit assembly would have been used, the light output must be as non-telecentric as the field rays coming from the telescope. The non-telecentricity is compensated in the spectrograph by a large quartz lens having about 650 mm in diameter. Due to the fact the lens is fixed in place the IFU output has to feed the collimator with the same lack of telecentricity as the telescope itself. To achieve these requirements, a slit curvature has to be introduced. Additionally the output lenslet arrays are used for beam steering to maintain the telescope's non-telecentricity.

\begin{figure}[h]
\begin{center}
\begin{tabular}{cc}
\includegraphics[height=6cm]{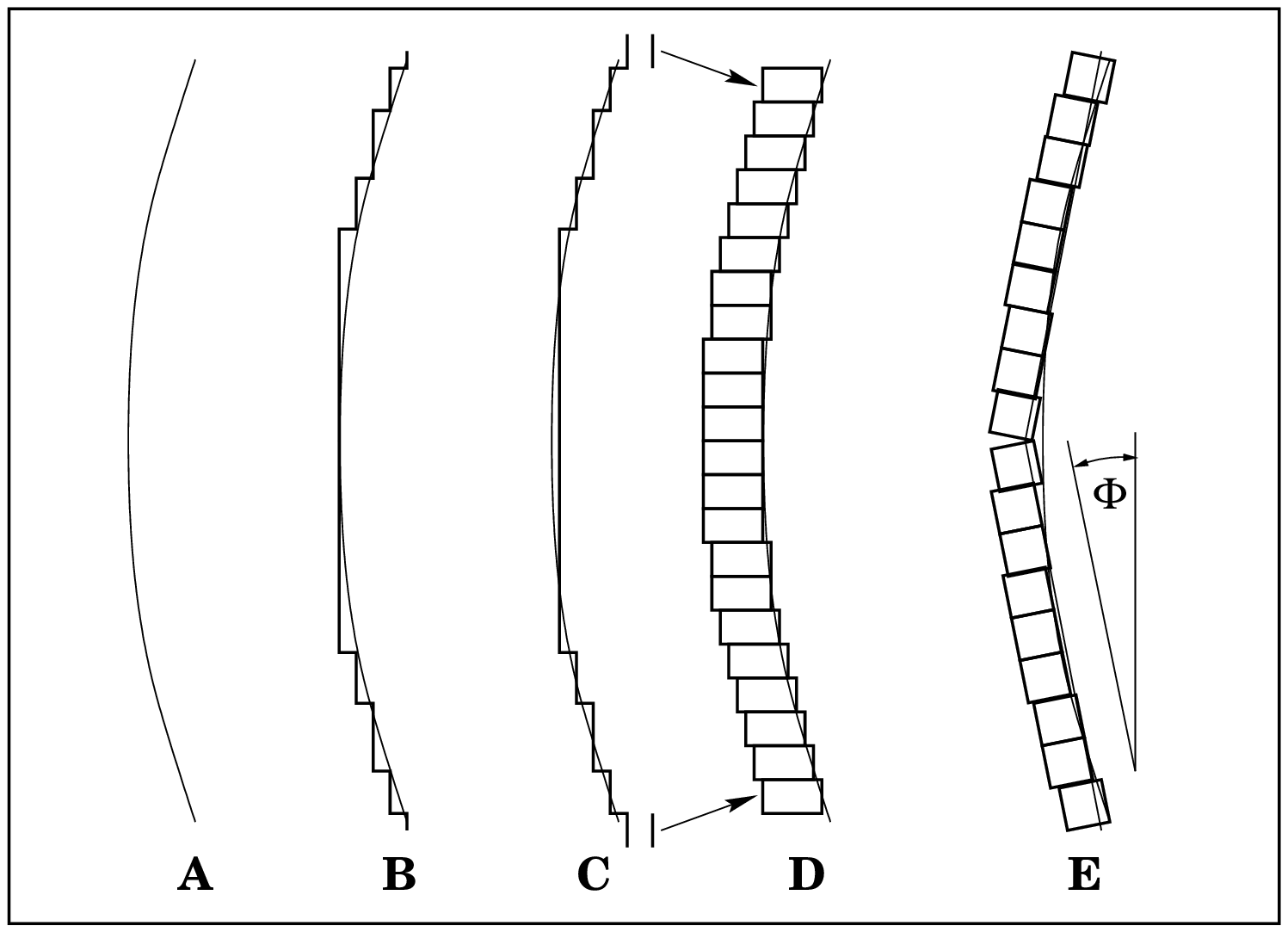}
\includegraphics[height=6cm]{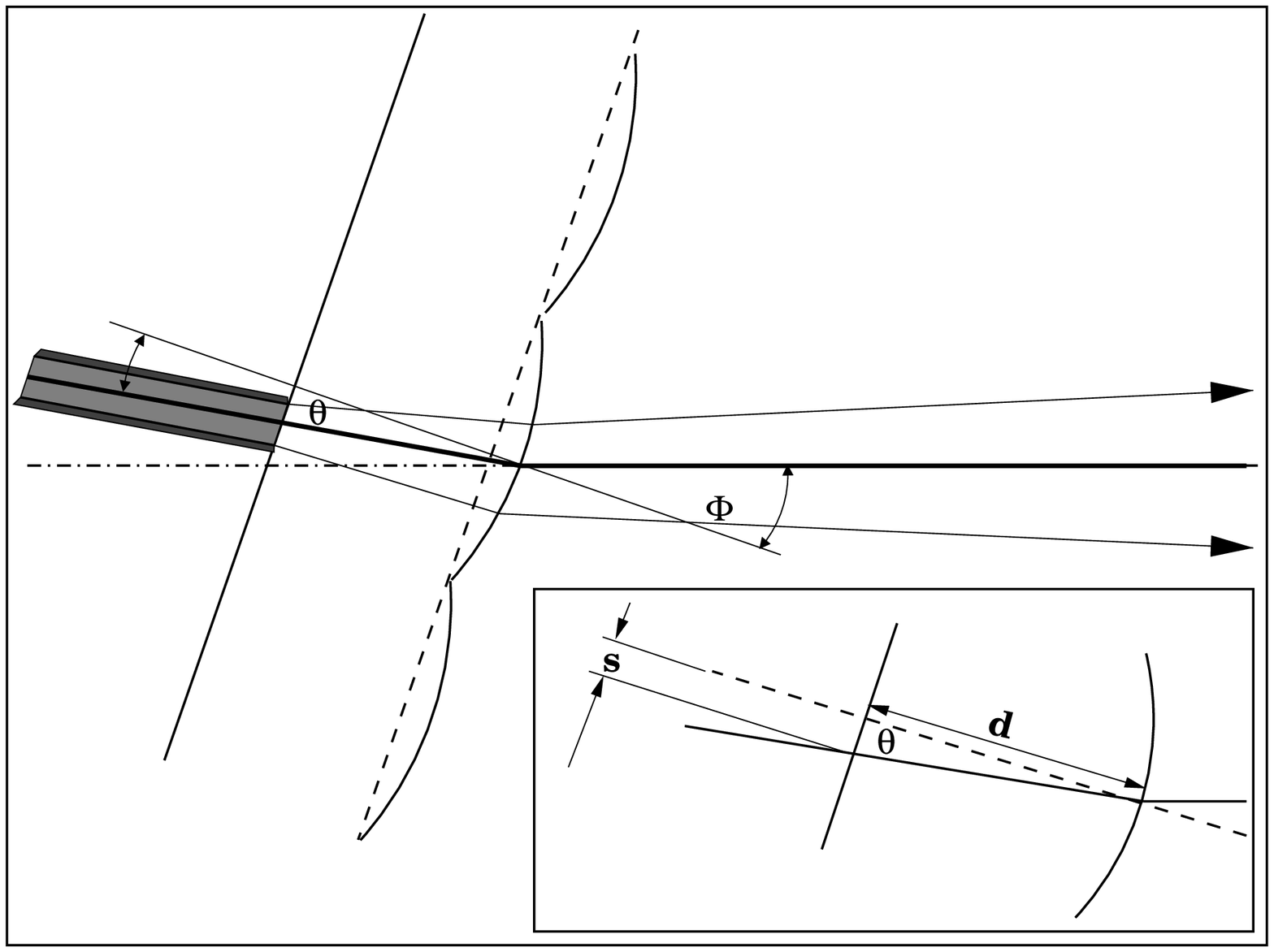}
\end{tabular}
\end{center}
\caption{Left: Illustration of approach adopted for approximation by a staircase-like pattern, E showing the solution of two tilted sub-slits that has been chosen. Right: Beam steering situation for the tilted fibers and offset micro lenses.\label{fig:slitgeometry}}
\end{figure}

As schematically shown in figure \ref{fig:slitgeometry} (left), the curved slit is realized by grouping the 2000 fibers into 40 blocks of 50 fibers each. A maximum defocus of $\pm$ 1 mm was allowed to allow a feasible group size. While the steps A to D in this picture show the process of block calculation, E shows the chosen design. The slit is divided into two halves, which are tilted by about 6 degrees to the normal of the optical axis. The resulting tilt of the fibers relative to the optical axis was compensated by a lateral shift of the output lenslet arrays. For production simplification, all blocks have been ground to a single angle, giving rise to a small variation of efficiency along the slit. The lateral offset change due to variation of the non-telecentric angle along a single block is less than 2 $\mu$m and thus negligible. Figure \ref{fig:slitgeometry}, right shows the situation for the slit approximated by two staircase-like patterns of output blocks as seen for case E.

\subsection{Mapping}
\label{sect:mapping}

\begin{figure}[h]
\begin{center}
\begin{tabular}{c}
\includegraphics[width=14cm]{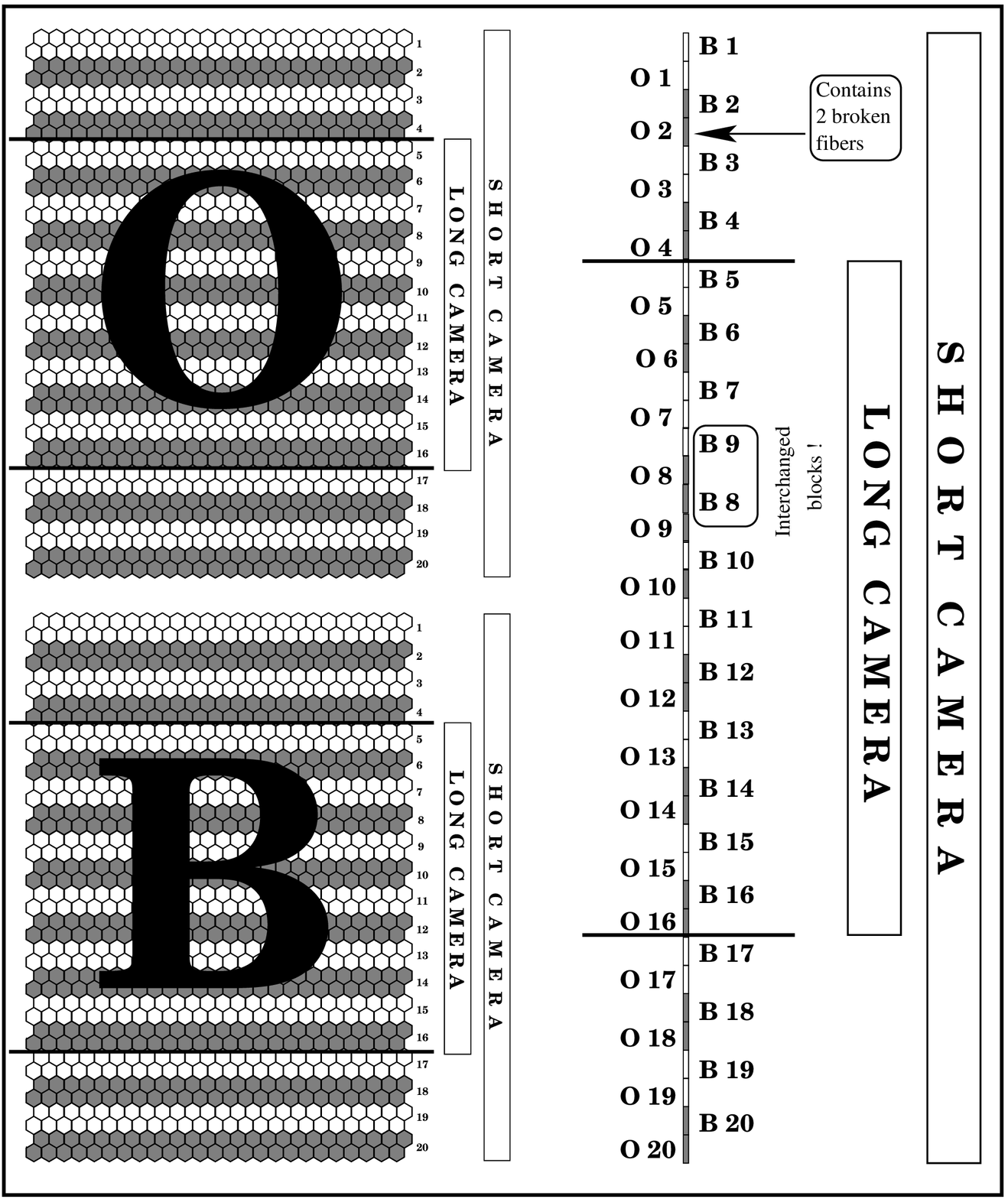}
\end{tabular}
\end{center}
\caption{Final mapping of the IMACS IFU. Adjacent blocks of the object field (labelled O) will be separated by a block of the background field (labelled B). \label{fig:mapping}}
\end{figure}

The mapping of the two sky fields was chosen in a way to assure that central elements of the fields occur near the slit center. Hence the central part of the slit corresponds to the inner parts of the fields, and while sky coverage is lost using the long camera, the locations of the field centers on the sky are not changed. Furthermore the blocks along the slit come alternately from the object and the background fields to avoid large changes in the spectrograph behaviour when the background is to be subtracted. The wings of the point spread function can be traced to a further extent between adjacent blocks, and also the curvature of the spectra can be traced using these gaps. The replacement of the double aperture mask at the IFU input by a single one allows investigation of a single field output, and the behaviour of the far outer wings of the PSF in cross dispersion direction can be studied in that way. The final mapping, including some minor errors of manufacture, is shown in figure \ref{fig:mapping}. Also the accessible ranges for the long and short cameras are pointed out in that figure, showing the restrictions of the long camera that images the inner 24 fiber blocks only out of the slit being 40 fiber blocks long. This gives rise to a smaller field of view for both fields, while the field centers are unchanged. The manufacture errors mentioned are two lost fibers in block O2 and two blocks (B8 and B9) accidently interchanged. This interchange has no disadvantage as long as it is accounted for in the data reduction.


\section{MECHANICAL DESIGN} 
\label{sect:mechanicaldesign}

\begin{figure}[h]
\begin{center}
\begin{tabular}{cc}
\includegraphics[height=8cm,angle=270]{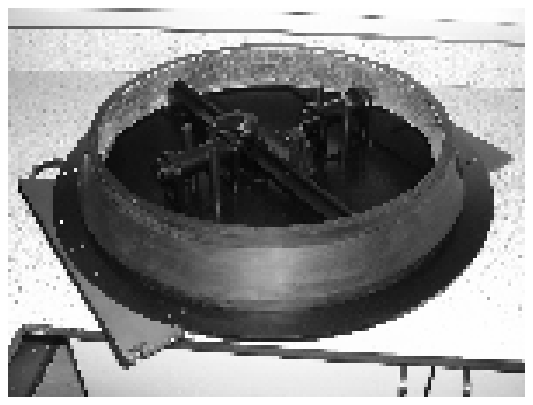}
\includegraphics[height=8cm,angle=270]{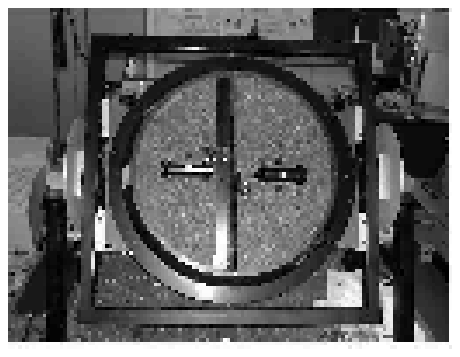}
\end{tabular}
\end{center}
\caption{The IFU housing, left holders on base plate, right the same after the foam layer has been installed. The frame on the right side was used for installing and alignment of the IFU. \label{fig:housingpictures}}
\end{figure}

The main issues of the mechanical design were to maintain overall alignment while fitting into the spatial and mass envelopes. A base plate was used as reference for building up the whole system, while the cover stiffens the plate from the edge.
\subsection{Base plate}
The base plate consists of milled aluminium tooling plate. To save mass, the plate thickness has pockets milled down to 0.5 mm where the original stiffness of 6.35 mm was not necessary. Two wings are bolted to the base plate to interface with the slit mask server. The stiffness is strongly increased by use of the housing as stabilizer. The structure has been FEA analyzed to assure that the flexure is within the tolerances for all gravitational situations. While the IFU moves between spring loaded rollers, it is hooked into position by a single lock, making a detailled analysis necessary. As result of the calculations the maximum flexure was about 5 $\mu$m and easily within specification.

\subsection{Optical mounts}
The optical mounts have been made out of aluminium. Two holders hold each of the two fore-optics barrels. The distances of the elements have been calculated and refined after the doubletts have been manufactured and measured. The pickoff mirrors are glued to spring-loaded rectangular adjusters, allowing a comfortable alignment in perpendicular directions. After the laboratory alignment had been finished, the adjusting screws have been locked using nuts and sealed with protecting glue.
\subsection{Output slit comb}
Because the comb stability is essential for the overall performance, this unit consists of a quite massive CNC machined aluminium bridge. Pointing grooves assure that the tilt of the output blocks is maintained by pressing them to a reference surface before cementing them into position. A gauge tool assured that the distance to the curved focal plane is maintained during the cementing process.
\subsection{Housing}
The housing was made out of fibre glass material to reduce mass. It has been equipped with internal threads to be connected to the base plate using bolts. The two object aperture holes can be covered with a shutter for periods when the IFU is not in use. On two sides there are pockets for silica gel packets to keep the inner IFU parts dry. The indicating silica gel can be seen and accessed from outside to replace it if necessary.


\section{COMMISSIONING AND FIRST EXPERIENCES}
\label{sect:commissioning}
The IMACS-IFU was shipped to Las Campanas in September 2003. Before that, some initial tests were made to assure the efficiency is within the expectation. For this a photodiode has been used in combination with an f/11 input beam. The light emerging from four blocks randomly chosen indicated a throughput between 66 and 74 \%, with a mean value of 70 \%. While this figure represents the value theoretically expected, its value had not to be overinterpreted due to the coarse sampling and the simplified setup involved.

\subsection{Commissioning}
Soon after receipt of the IFU first tests at the telescope have been done. Apart from some minor mechanical interface problems the IFU performed well and the slit mask server had no problem handling the 10 kg load. One problem was that the undispersed slit image just fell into a gap of the CCD mosaic. To avoid this, the wings that adapt the round base plate to the rails of the mask changer had to relocated slightly.

\subsection{PSF shapes and throughput}

\begin{figure}[h]
\begin{center}
\begin{tabular}{c}
\includegraphics[height=8.2cm]{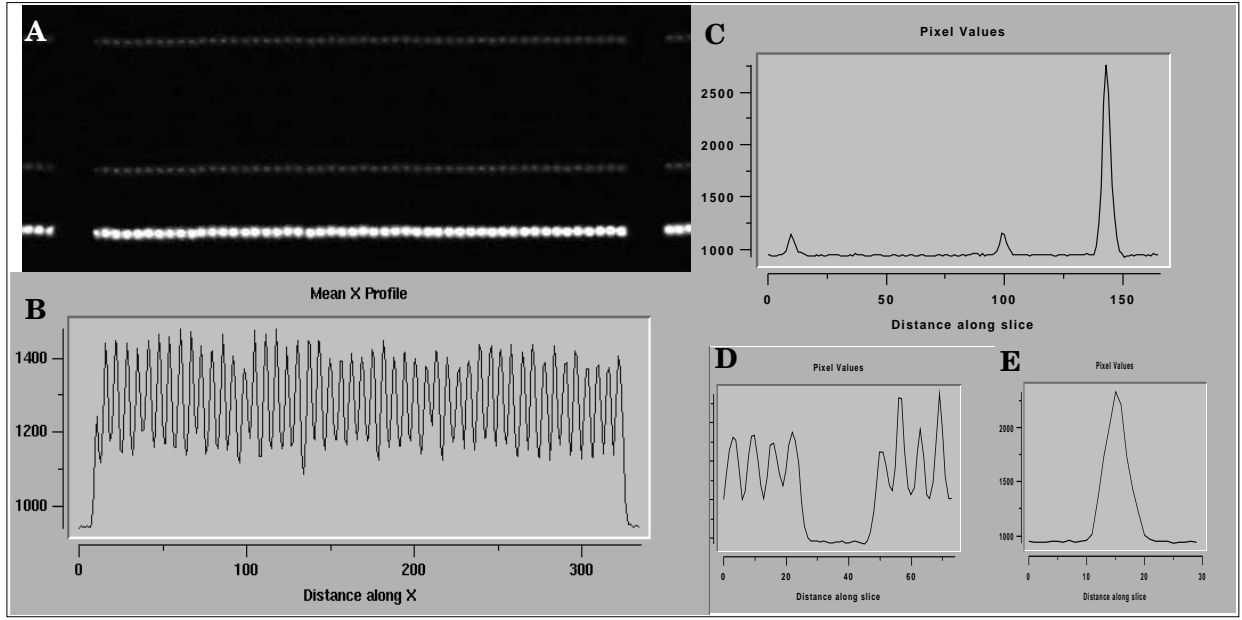}
\end{tabular}
\end{center}
\caption{Arc image as seen through the IMACS IFU. A: Linear scaled image, B: Profile plot along one block, C: Plot along 3 emission lines, D: Cross-dispersion plot between two blocks, E: A single line along dispersion. \label{fig:arc_analysis}}
\vspace{5mm}
\end{figure}

\begin{figure}[h]
\begin{center}
\begin{tabular}{c}
\includegraphics[height=12cm]{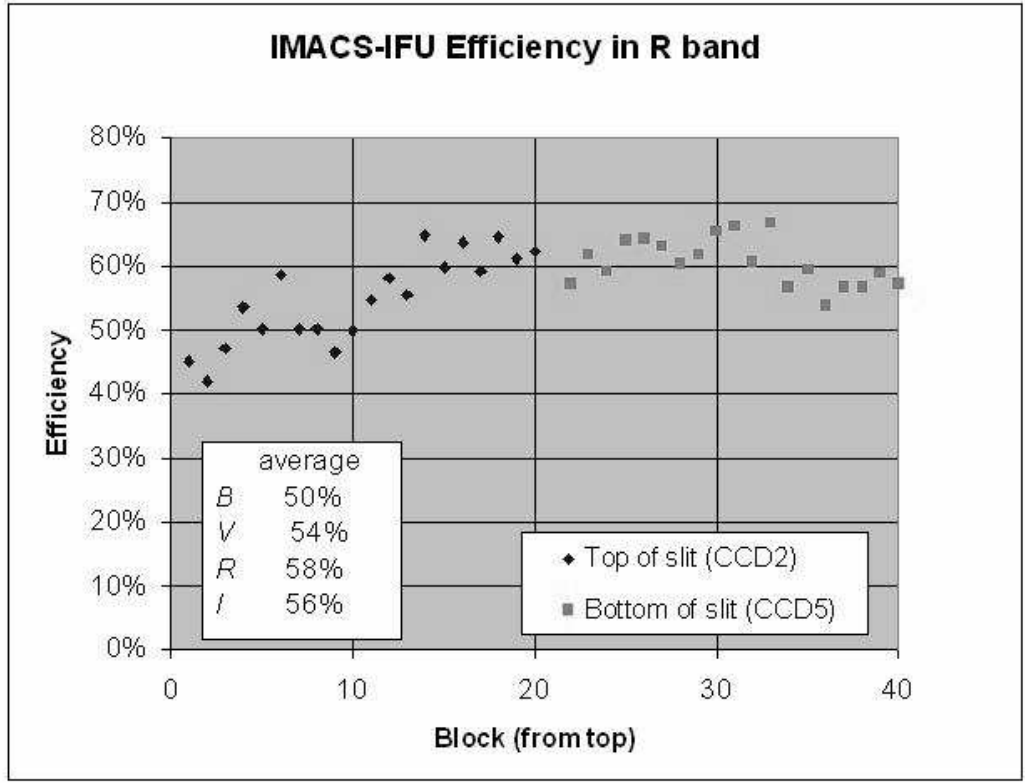}
\end{tabular}
\end{center}
\caption{Efficiency of the IMACS IFU at the telescope, using white light flat field illumination and broad band filters. \label{fig:imacs_ifu_throughput_nov03_bw}}
\vspace{5mm}
\end{figure}

The calibration lamp images obtained revealed the results of figure \ref{fig:arc_analysis} after raw focusing. The spectra of adjacent fibers are easily to distinguish, with overlap at about the FWHM of each profile. The background between adjacent fiber blocks is quickly reached, which indicates low levels of scattered light. The spacing represents the expected 3.54 pixel per spectrum for the short camera being used. In spatial and wavelength direction the point spread functions are similar. The throughput has been measured using an artifical continuum light source in the intermediate pupil near the Gregorian secondary mirror, approximating an even illumination of the focal plane using a flat screen. Figure \ref{fig:imacs_ifu_throughput_nov03_bw} shows the throughput as a mean value of far different broadband filter passbands. The figure is split into two half sections to indicate that each half is projected onto a different CCD chip which may have different gains and quantum efficiency. While the specification of an efficency greater or equal to 50 \% is reached on the right end, the blocks at the outer left do not reach this spec always. The output microlens arrays were examined by microscope before assembly to assure that the best ones are used at the slit center. The left half of the slit contains one micro lens array more with a slightly poorer performance than the right side. Also the left side holds the block with the two dead fibers, explaining a poorer efficiency of this block (number 3 in figure \ref{fig:imacs_ifu_throughput_nov03_bw}).

\subsection{Mechanical stability}

As far as it could be seen on the first runs, there has been no obvious problem with the mechanical rigidity of the IFU. The opics inside kept alignment over the transport to Las Campanas, and no obvious flexure was visible during exposures of several ten minutes, neither inside the IFU nor in the mask holder mechanism of the spectrograph. However furthermore investigations and experiences have to be done here to quantify this behaviour.


\section{CONCLUSIONS}
Although some slit blocks are slightly less efficient than specified, the IMACS-IFU overall performance is acceptable. The results achieved show that an fiber-lenslet based IFU is possible even with a very long and highly curved slit necessary to deal with the extreme field curvature and non-telecentricity of the telescope. Also it shows that an IFU of that size can be kept modular and easy to handle. The furthermore potential using beam-switching between the two similar sized fields is still to be exploited.


\acknowledgments     

Graham Murray and David Robertson, both of our group, are gratefully acknowledged for their collaboration. Graham was sharing ideas and experiences of former IFUs, while David Robertson is acknowledged for doing higher level project management.



\end{document}